\documentclass[12pt]{article}
\usepackage{amssymb,epsf,color}
\def\lsim{\mathrel{\rlap {\raise.5ex\hbox{$ < $}}
{\lower.5ex\hbox{$\sim$}}}}
\def\gsim{\mathrel{\rlap {\raise.5ex\hbox{$ > $}}
{\lower.5ex\hbox{$\sim$}}}} 
\def\sqr#1#2{{\vcenter{\vbox{\hrule height.#2pt

        \hbox{\vrule width.#2pt height#1pt \kern#1pt

           \vrule width.#2pt}

        \hrule height.#2pt}}}}

\def\lsim{{\displaystyle
{{\raise-8pt\hbox{$ <$}}
\atop{\raise5pt\hbox{$\sim$}}}}}
\def\gsim{{\displaystyle
{{\raise-8pt\hbox{$ >$}}
\atop{\raise5pt\hbox{$\sim$}}}}}
\def\slsim{{\displaystyle
{{\raise-8pt\hbox{$\scriptstyle <$}}
\atop{\raise5pt\hbox{$\scriptstyle \sim$}}}}}
\def\sgsim{{\displaystyle
{{\raise-8pt\hbox{$\scriptstyle  >$}}

\atop{\raise5pt\hbox{$\scriptstyle \sim$}}}}}

\newskip\humongous \humongous=0pt plus 1000pt minus 1000pt

\newcommand{\sumpf}[0]{\sum_{(H^{\rm f},G^{\rm f})}\! \! \! \!
{\raise
4pt
\hbox{$'$}}\,}

\newcommand{\sump}[0]{\sum_{(H,G)}\! \! {\raise 4pt \hbox{$'$}}\,}

\def\bs{\begin{subequations}}
\def\es{\end{subequations}}

\catcode`\@=11
\newcount\hour
\newcount\minute
\newtoks\amorpm

\hour=\time\divide\hour by60
\minute=\time{\multiply\hour by60 \global\advance\minute by-\hour}
\edef\standardtime{{\ifnum\hour<12 \global\amorpm={am}%
        \else\global\amorpm={pm}\advance\hour by-12 \fi

        \ifnum\hour=0 \hour=12 \fi
        \number\hour:\ifnum\minute<10 0\fi\number\minute\the\amorpm}}
\edef\militarytime{\number\hour:\ifnum\minute<10 0\fi\number\minute}
\def\draftlabel#1{{\@bsphack\if@filesw {\let\thepage\relax
   \xdef\@gtempa{\write\@auxout{\string
      \newlabel{#1}{{\@currentlabel}{\thepage}}}}}\@gtempa
   \if@nobreak \ifvmode\nobreak\fi\fi\fi\@esphack}
        \gdef\@eqnlabel{#1}}
\def\@eqnlabel{}
\def\@vacuum{}
\def\draftmarginnote#1{\marginpar{\raggedright\scriptsize\tt#1}}
\def\draft{\oddsidemargin -.2truein
        \def\@oddfoot{\sl preliminary draft \hfil
        \rm\thepage\hfil\sl\today\quad\militarytime}
        \let\@evenfoot\@oddfoot \overfullrule 3pt
        \let\label=\draftlabel
        \let\marginnote=\draftmarginnote
   \def\@eqnnum{(\theequation)\rlap{\kern\marginparsep\tt\@eqnlabel}%
\global\let\@eqnlabel\@vacuum}  }
\def\subequations{\refstepcounter{equation}%
  \edef\@savedequation{\the\c@equation}%
  \@stequation=\expandafter{\theequation}
  \edef\@savedtheequation{\the\@stequation}
  \edef\oldtheequation{\theequation}%
  \setcounter{equation}{0}%
  \def\theequation{\oldtheequation\alph{equation}}}

\def\endsubequations{\setcounter{equation}{\@savedequation}%
  \@stequation=\expandafter{\@savedtheequation}%
  \edef\theequation{\the\@stequation}\global\@ignoretrue
  \vspace*{-12pt} \\}

\def\bs{\begin{subequations}}
\def\es{\end{subequations}}
\relax
\def\thefootnote{\fnsymbol{footnote}}
\def\be{\begin{equation}}
\def\ee{\end{equation}}
\def\ba{\begin{eqnarray}}
\def\ea{\end{eqnarray}}

\def\ee{\end{equation}}
\def\bea{\begin{eqnarray}}
\def\eea{\end{eqnarray}}
\def\nn{\nonumber}

\newcommand{\uarrw}[0]{\mathrel{
{\raise.5ex\vbox{\hrule width 1cm}\hskip-6pt\rightarrow}}}
\def\thebibliography#1{%
\vskip 0.5cm \centerline{\bf References}
\list{%
[\arabic{enumi}]}{\settowidth\labelwidth{[#1]}
\leftmargin\labelwidth
\advance\leftmargin\labelsep
\usecounter{enumi}}
\def\newblock{\hskip .11em plus .33em minus .07em}
\sloppy\clubpenalty4000\widowpenalty4000
\sfcode`\.=1000\relax}

\renewcommand{\theequation}{\arabic{section}.\arabic{equation}}

\renewcommand{\section}{\setcounter{equation}{0}\@startsection%
{section}{1}{0mm}{-\baselineskip}{0.5\baselineskip}%
{\normalfont\normalsize\bfseries}}

\renewcommand{\subsection}{\@startsection%
{subsection}{2}{0mm}{-\baselineskip}{0.5\baselineskip}%
{\normalfont\normalsize\slshape}}

\renewcommand{\subsubsection}{\@startsection%
{subsubsection}{2}{0mm}{-\baselineskip}{0.5\baselineskip}%
{\normalfont\normalsize\slshape}}

\begin{document}
%
%
\renewcommand{\theequation}{\arabic{section}.\arabic{equation}}
\begin{titlepage}
\begin{flushright}
\end{flushright}
\begin{centering}
\vspace{1.0in}
\boldmath

{ \large \bf Relativity as classical limit in a combinatorial scenario \\
\bf }

\unboldmath
\vspace{2.5 cm}

{\bf Andrea Gregori}$^{\dagger}$ \\
\medskip
\vspace{2.2cm}
{\bf Abstract} \\
\end{centering} 
\vspace{.2in}
We discuss how the finiteness and universality of the speed of light arise in 
the theoretical framework introduced in \cite{assiom}, and derive generalized
coordinate transformations, that allow to investigate physical systems in a
non-classical regime.

\vspace{8cm}

\hrule width 6.7cm
\noindent
$^{\dagger}$e-mail: agregori@libero.it

\end{titlepage}
\newpage
\setcounter{footnote}{0}
\renewcommand{\thefootnote}{\arabic{footnote}}

\tableofcontents

\vspace{1.5cm}

\noindent

\section{Introduction}
\label{intro}

In Ref.~\cite{assiom} we introduced an 
entropy-weighted sum over all possible configurations $\psi$
of a discrete space
with a given amount of energy content $E \equiv N$, i.e. over all the
ways of distributing $N$ energy 
units along a space of unit cells. This had to serve as the generating function
for the (mean) values of any observable of a system that we call the
``universe'': 
\be
{\cal Z}_{E} \, = \, \int_E {\cal D} \psi \; {\rm e}^{{1 \over k} S} \, ,
\label{ZSintegral}
\ee
where $k$ is the Boltzmann constant, and
the entropy $S$ is intended in its statistical sense,
as the logarithm of the volume of occupation of a configuration $\psi(E)$
in the
phase space of all the configurations, 
$\left\{ \psi(E)  \right\}$~\footnote{In the following we will 
omit, wherever possible, namely, wherever the 
omission does not generate confusion,
any dimensional constant.}. The space coordinates, that take only discrete 
values multiples of a unit, are 
allowed to run over any number of dimensions and extension.
This unit had to
be viewed as a minimal length, and, in the limit to the continuum,
it has to be identified with the Planck length. 
In a loose sense, the entropy-weighted sum can be considered
a sum over all possible ``geometries'' \footnote{If we consider as a geometry
any possible distribution (assignment) of ``energy'' along discrete
space coordinates, that only for large numbers and in a limit can be 
approximated with the usual concept of continuous, differential geometry based
on the idea of dimensionless point.}, and as such contains all the 
information about the physics of the universe. 
The evolution of the universe, or in other words the time ordering, occurs
through inclusion of sets. Owing to the property:
\be
\left\{ \psi  \right\}_{E^{\prime}} \supseteq 
\left\{ \psi  \right\}_{E}   ~~~~~ {\rm if} ~~~~~
E^{\prime} \geq E \, ,
\ee  
the natural evolution is toward increasing total energy, that can also
be identified, via appropriate conversion of units, with a time, the ``age''
of the universe. Expression 
$\left\{ \psi  \right\}_{E^{\prime}} \supseteq 
\left\{ \psi  \right\}_{E}$ has to be intended in the sense that 
$\forall \psi_E \, \in \left\{ \psi  \right\}_{E}$, 
$\exists \, \psi_{E^{\prime}} \, \in \left\{ \psi  \right\}_{E^{\prime}}$ 
such that $\psi_E \subseteq \psi_{E^{\prime}}$~\footnote{The ``past'' 
is something of which the system conserves a memory. The configuration
it corresponds to belongs therefore also to 
the set of configurations at any subsequent time, 
as expressed by the inclusion relation of above.}.

From \ref{ZSintegral}
one can immediately see that the universe will be dominated by the 
configurations of maximal entropy. What is not at all obvious is that indeed
these are the ones with three space dimensions, and with a distribution of 
energy corresponding to a sphere with a radius/energy relation given
by the Schwarzschild black hole expression. Moreover, the fluctuations
in the mean energy value due to the contribution of the ``sea'' of non
dominant, in general not even geometrizable, configurations correspond to
the Heisenberg's time/energy Uncertainty Relation. We concluded therefore
in Ref.~\cite{assiom} that the functional \ref{ZSintegral} describes a quantum
scenario (or, more precisely, it embeds a quantum scenario). 

\vspace{1cm}

The quantum physical world as we experience it corresponds to the ``mean value
configuration'' of this scenario, and it turns out that the horizon of
this black hole-like universe expands at the speed of expansion of time,
by definition/choice of units = $c$ 
\footnote{As discussed in \cite{assiom} this
non-accelerated expansion nevertheless appears to be accelerated, as the 
effect of a time-dependence of masses and couplings, and therefore of 
the frequencies in the radiation/emission spectra.}.  
Here we will discuss how, in this universe, $c$ is also the
maximal speed of propagation of information. This leads to Relativity as a 
direct implication of this scenario. On 
the other hand, through this derivation we will also learn something
more about the limitations implicit in a description of physical phenomena
corresponding to the classical
relativistic limit, and gaining some
insight into the behaviour of physical systems beyond the domain of validity
of the theory of Relativity.

We will start by showing how the speed
of expansion of the radius of the average three-dimensional, black-hole-like 
universe is also the maximal speed of propagation of information 
(section~\ref{speedlight}). We say 
``information'', because, as also discussed in \cite{assiom}, in
the sum \ref{ZSintegral} are
also contained configurations with faster objects (tachyonic configurations),
but for these what is transferred cannot be considered something with
a (geometric) structure which is conserved during transportation: by 
``information'' we mean something that conserves its characteristics, 
i.e. carries, transfers, a message during its displacement.

We discuss then how different observers can relate their time measurements
(section~\ref{Lboost}). According to our approach, 
time has progressed as soon as something in the universe, which,
we remember it, \emph{includes} also the observer \footnote{We recall 
that in our theoretical approach there 
is no absolute universe in itself, as something independent on the relative
position of its observer: when we say that the universe has a certain age and 
the horizon a certain extension, this is the age and the extension as they
are measured by a particular observer.}, has changed.
If nothing changes, there is, by definition, no time progress.
In the framework underlying \ref{ZSintegral}, configurations
are identified by their energy distribution, i.e. by their
combinatorics of energy and space degrees of freedom. This means,
equivalently, through their symmetry group. As we work with discrete groups,
there are no two inequivalent configurations with the same symmetry group 
$\leftrightarrow$ volume of occupation in the phase space $\leftrightarrow$
entropy. In particular, this holds also for the dominant configuration
at any time. It is therefore possible to correlate the time flow to
the entropy change of the dominant configuration. Similarly 
the perception of time in any subsystem of the universe
must be related to the entropy of the (subset of the) dominant configuration
corresponding to that subsystem.
This allows us to rephrase the problem of time transformation between
different frames into a problem of transformation of entropies.  
In this work the Lorentz boosts are derived via entropy arguments. 
This opens up new 
perspectives, eventually allowing to determine the metric of space-time
around any point in the universe directly from the formulation in terms
of entropy. That is, beyond classical General Relativity, in a pure
``quantum gravity'' regime.

We pass then, in the following section~\ref{gtrans}, to the
general coordinate transformation, and therefore also to the metric in 
itself. All the relations we discuss involve only the time component.
Time is in fact the ground quantity in this framework,
as it is directly related to the fundamental quantities of
this scenario, namely energy and entropy.
Space lengths are derived quantities: lengths are defined through measurements
of extensions and displacements that take place during a time interval, and
therefore are related to time and to the speed of the transferred information.
On the other hand, this is true also for the traditional derivation
of the Lorentz boosts: the Lorentz space boost is
obtained by comparing measurements of length made by counting the time
needed for light to travel from one edge to the other of a space segment.
In a more involved way, it is in principle possible also in our framework
to obtain the full coordinate transformations, something we will however not
do here, for the sake of shortness and simplicity, as it would add no further 
fundamental ideas and deeper insight into the problem.

Raising the Einstein's equations to entropy-dependent expressions 
provides us with a more general expression, valid also in a
quantum scenario and for cases in which no geometry in a classical sense
can be identified. At the end of the work (section~\ref{beyondc}) 
we briefly discuss time coordinate transformations
in the case of complex quantum systems 
in a quantum/relativistic regime.

\section{From the speed of expansion of the universe to a maximal speed for 
the propagation of information}
\label{speedlight}

As discussed in Ref.~\cite{assiom}, the universe arising from the superposition
of all possible energy distributions (= configurations) 
at time $E = N$ 
is predominantly a three dimensional one with an energy distribution
corresponding to the geometry of a three-sphere with radius $\sim N$. 
Such a three dimensional universe, of radius $\sim N$, at time $N$, 
with total energy
also $N$, behaves like a black hole expanding at speed $1$ (we can introduce
a factor of conversion from time to space, $c$, and say that, by choice
of units, we set the speed of expansion to be $c = 1$).  
Here we want to see how this scenario implies that this is also the 
maximal speed for propagation of information within the three dimensional
universe (i.e., inside the black hole).
It is important to stress that all this refers only to the average universe, 
because only in this sense we can say that the universe is three dimensional:
the sum \ref{ZSintegral} contains in fact also configurations in which higher 
speeds are 
allowed (we may call them ``tachyonic'' configurations), along with
configurations in which it is not even clear what is the meaning of
speed of propagating information in itself, as there is no recognizable 
information at all, at least in the sense we usually intend it.

Indeed, when we say we get information about, say, the motion of a particle,
or a photon,
we mean to speak of a non-dispersive wave packet, so that we can say
we observe a particle, or photon, that remains particle, or photon,
along its motion \footnote{Like a particle, also a physical photon, 
or any other field,
is not a pure plane wave but something localized, therefore a 
superposition of waves, a wave packet.}.
Let's consider the simplified case of a universe
at time $N$ containing only one such a wave packet \footnote{We may think to
concentrate onto only a portion of the universe, where only such a wave
packet is present.}, as illustrated in 
figure~\ref{gridN}, where it is represented by the shadowed cells, and
the space is reduced to two dimensions.
\be
\epsfxsize=6cm
\epsfbox{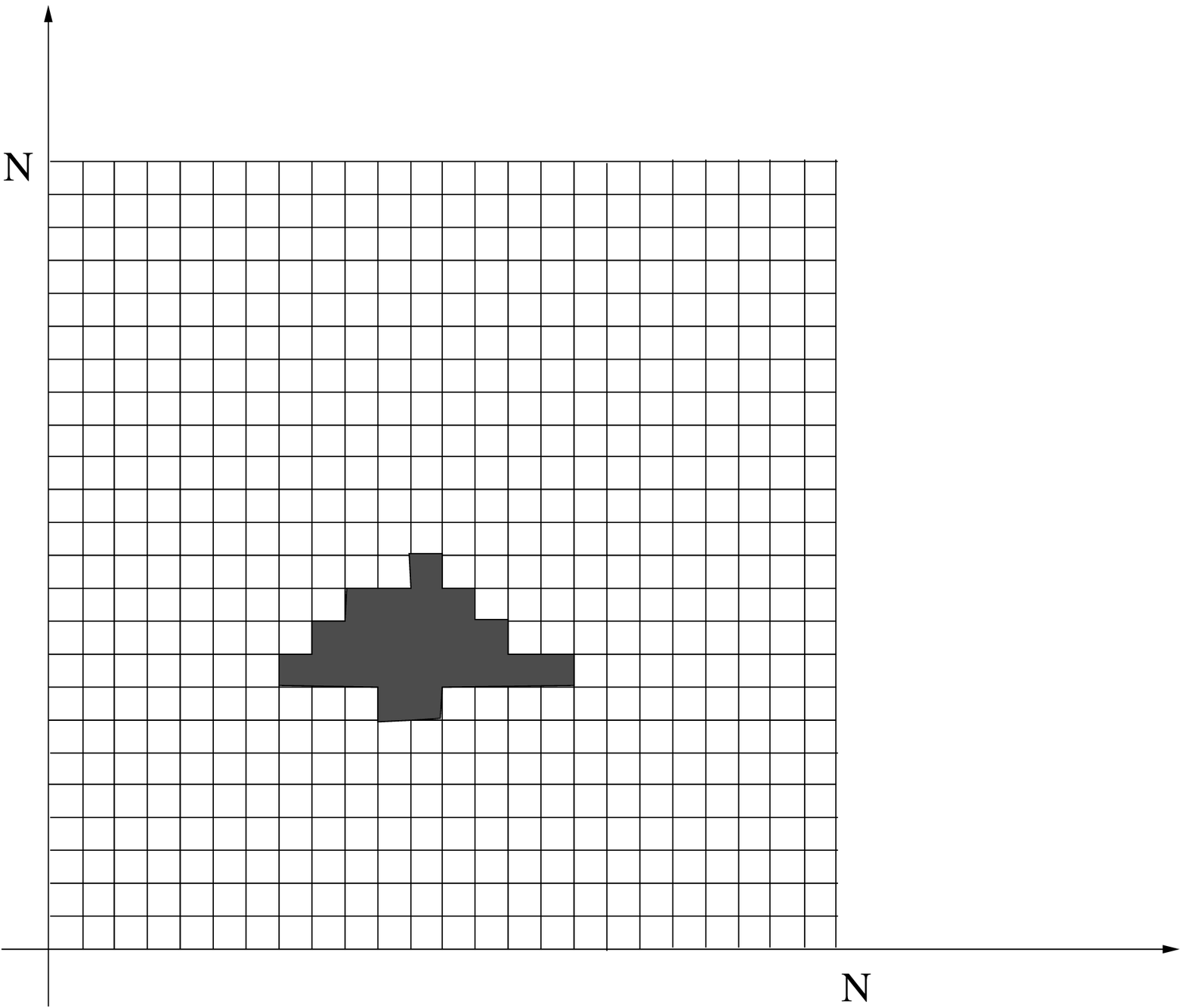}
\label{gridN}
\ee
Consider now the evolution at the subsequent instant of time,
namely after having progressed by a unit of time. 
Adding one point, $N \to N+1$, does 
produce an average geometry of a three 
sphere of radius $N+1$ instead of $N$. In the average, it is therefore like 
having added $4 \pi N^2$ ``points'', or unit cells. 
Remember that we work always with an 
infinite number of cells in an unspecified number of dimensions; when we talk 
of universe in three dimensions within a region of a certain radius, 
we just talk of the average geometry, 
in the sense explained in Ref. \cite{assiom}.
Let's suppose the position of the
wave packet jumps by steps (two cells) back, 
as illustrated in figure~\ref{gridN+1}. Namely, as time, and consequently also
the radius of the universe, progresses by one
unit, the packet moves at higher speed, jumping by two units: 
\be
\epsfxsize=6cm
\epsfbox{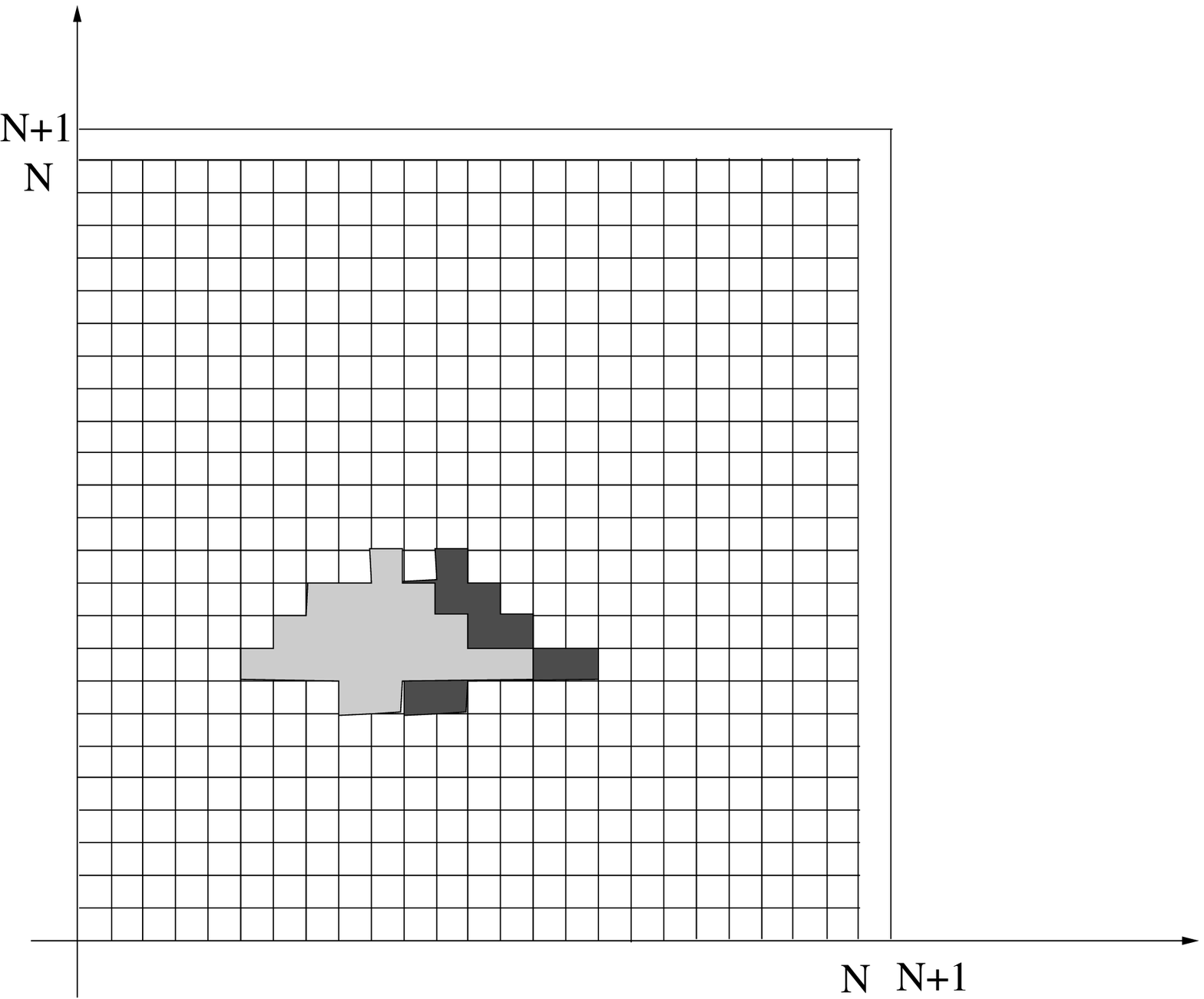}
\label{gridN+1}
\ee
Consider now the case in which the packet jumps by just one unit,
as in figure~\ref{gridN+1bis} here below: 
\be
\epsfxsize=6cm
\epsfbox{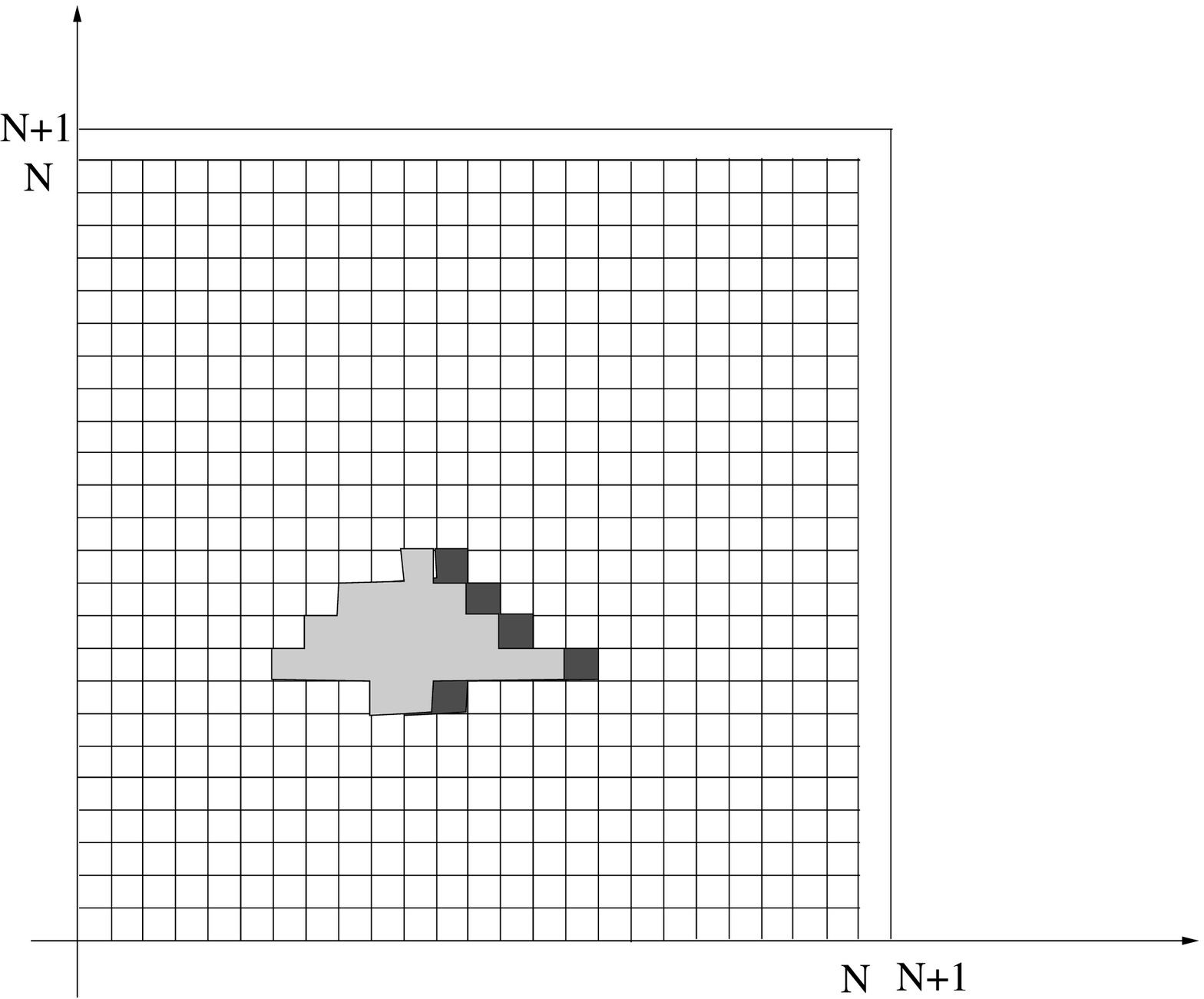}
\label{gridN+1bis}
\ee
The entropy of this latter configuration, intermediate between the first and 
second one, cannot be very different from the one of the second configuration,
figure~\ref{gridN+1}, in which the packet
jumps by two steps, because that was supposed to be the dominant configuration
at time $N+1$, and therefore the one of maximal entropy. Indeed, 
by ``continuity'' it must interpolate between step 2 and the configuration
at time $N$, that was also supposed to be a configuration of maximal
entropy. Therefore, the actual
appearance of the universe at time $N+1$ must be somehow a superposition
of the configurations 2 and 3, thereby contradicting our hypothesis that
the wave packet is non-dispersive~\footnote{If it was dispersive,
it would be something like a particle that, during its motion, ``dissolves'', 
and therefore we cannot anymore trace as a particle. It would be just
a ``vacuum fluctuation'' without true motion, something that does not carry
any information.}. Therefore, the wave packet cannot jump by two steps, 
and we conclude that the maximal speed allowed is that of expansion of the 
radius of the universe itself, namely, $c$.

As discussed in Refs.~\cite{assiom,spi}, 
\ref{ZSintegral} leads to the ordinary physics and the universe as we know it,
with the correct content of particles, fields and interactions. 
According to this theoretical framework,  
the reason why we have a universal bound on the speed of light
is therefore that light carries what we call classical information. 
Information about whatever kind of event tells 
about a change of average entropy of the observed system, 
of the observer, and what surrounds and connects them too. 
The rate of transfer/propagation
of information is therefore strictly related to the rate of variation of
entropy. Variation of entropy is what gives the measure of time progress
in the universe. Any vector of information that ``jumps'' steps of the
evolution of the universe, going faster than its rate of entropy variation, 
becomes therefore dispersive, looses information
during its propagation. Light must therefore propagate at most at the rate of
expansion of space-time (i.e. of the universe itself). Namely, at the rate of  
the space/time conversion, $c$.

\section{The Lorentz boost}
\label{Lboost}

Let's now consider systems that can be identified as ``massive particles'',
i.e. localizable and that exist also ``at rest'', therefore travelling 
at speeds always lower than $c$. 
Since the phase space has a multiplicative structure, 
and entropy is the logarithm of 
the volume of occupation in this space, it is possible to separate for each 
such a system the entropy into the sum of an internal, ``rest'' entropy, and
an external, ``kinetic'' entropy. The first one refers to the structure
of the system in itself, that can be a point-like particle or an entire
laboratory \footnote{In our approach, there does not exist a strictly
point-like object. A point-like particle is an extended object of which
we neglect the geometric structure.}. 
The second one refers to the relation/interaction of this
system with the environment, the external world: its motion, the
accelerations and external forces it experiences, etc.      

\begin{figure}
\centerline{
\epsfxsize=7cm
\epsfbox{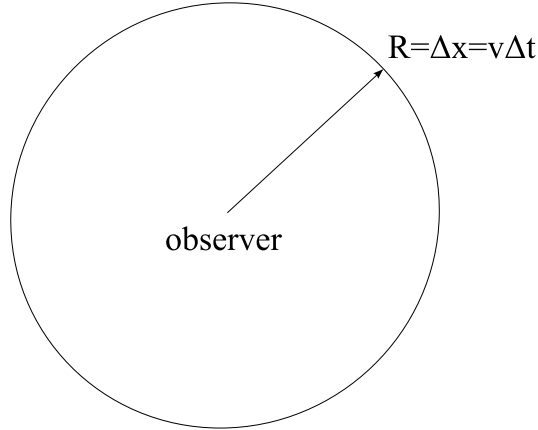}
}
\vspace{0.3cm}
\caption{During a time $\Delta t$, the pure motion ``creates'' a universe
with an horizon 
at distance $\Delta x = v \Delta t$ from the observer. As seen from the
rest frame, this part of the physical system does not exist. The ``classical''
entropy of this region is given by the one of its dominant configuration, i.e.
it corresponds to the entropy of a black hole of radius $\Delta x$.} 
\label{v-universe}
\end{figure}

Let us for a moment abstract from the fact that the actual
configuration of the universe implied by \ref{ZSintegral} at any time 
describes a curved space. In other words, let's neglect the so called
``cosmological term''. This approximation can make sense at large $N$,
as is the case of the present-day physics, a fact that historically
allowed to introduce special relativity and Lorentz boosts before addressing 
the problem of the cosmological constant.
Let us also assume we can just focus our attention on two observers sitting
on two \underline{inertial} frames, $A$ and $A^{\prime}$, 
moving at relative speed $v$, neglecting everything else \footnote{In our 
theoretical framework, there is no ``external observer'': \ref{ZSintegral} 
describes a universe ``on shell'', the totality of the physical world.}.  
For what above said, $v < 1$. 
An experiment is the measurement of some event that, owing to the
fact that happening of something means changing of entropy and therefore
is equivalent to a time progress \footnote{See Ref.~\cite{assiom}.}, 
gives us the perception of having taken place during
a certain interval of time. Let us
consider an experiment, i.e. the detection of some event, taking place in
the co-moving frame of $A^{\prime}$, as reported by both
the observer at rest in $A$,
and the one at rest in $A^{\prime}$ (from now on we will
indicate with $A$, and $A^{\prime}$, indifferently the frame as well as
the respective observer). Let's assume
we can neglect the space distance separating the two observers, or suppose
there is no distance between them \footnote{In our scenario, 
huge (=cosmological) distances
have effect on the measurement of masses and couplings.}. 
For what above said, such a detection amounts in observing
the increase of entropy corresponding to the occurring 
of the event, as seen from $A$, 
and from $A^{\prime}$ itself. Since we are talking of the same event, the
\emph{overall} change of entropy will be the same for both 
$A$ and $A^{\prime}$.
One would think there is an ``absolute'' time
interval, related to the evolution of the universe corresponding to the
change of entropy due to the event under consideration. However, 
the story is rather different as soon as we consider \emph{time measurements}
of this event, as reported by the two observers, $A$ and $A^{\prime}$. 
The reason is that the two observers will in general attribute
in a different way what amount of entropy change has to be considered
a change of entropy of the ``internal'' system, and which amount
refers to an ``external'' change. Proper time measurements have to do with
the \emph{internal} change of entropy. 
For instance, consider the entropy of
all the configurations contributing to form, say, a clock. The part of phase
space describing the uniform motion of this clock will not be taken into
account by an observer moving together with the clock, as it will not even
be measurable. This part will however be considered by the other observer.    
Therefore, when reporting measurements of time intervals made by 
two clocks, one co-moving with $A$, and one seen
by $A$ to be at rest in $A^{\prime}$, owing to a different way of
attributing elements within the configurations building up the system,
between ``internal'' and 
``external'', we will have in general two different time
measurements.  
Let us indicate with $\Delta S$ the change of entropy as it is observed
by $A$. We can write:
\ba
\Delta S \, (\equiv \Delta S(A) ) 
& = & \Delta S ({\rm internal}\, = \, {\rm at \, rest}) 
\, + \, \Delta S ({\rm external})  \label{DeltaSS} \\
&& \nn \\
& = & \Delta S (A^{\prime}) \, + \, 
\Delta S_{\rm Kinetic}(A) \, ,  
\label{DeltaSSk}
\ea
with the identifications 
$\Delta S ({\rm internal}
\, = \, {\rm at \, rest}) \equiv \Delta S (A^{\prime})$ and
$\Delta S ({\rm external}) \equiv \Delta S_{\rm Kinetic}(A)$.
In Ref.~\cite{assiom} we discussed how the entropy of a three sphere is 
proportional to $N^2 = E^2$. This is therefore also the entropy of 
the average, classical universe, that in the continuum limit, via the 
identification of total energy with time, can be written as: 
\be
S \; \propto \; \left( c {\cal T} \right)^2 \, ,
\label{ScT2}
\ee
where ${\cal T}$ is the age of the universe.
This relation agrees with the Hawking's expression of the entropy of a 
black hole of radius $r = c {\cal T}$, 
as indeed the universe in our theoretical framework is.
It is not necessary to write explicitly  the proportionality constant in 
(\ref{ScT2}),
because we are eventually interested only in ratios of entropies.
During the time of an event, $\Delta t$, the age of the universe passes
from ${\cal T}$ to ${\cal T} + \Delta t$, and
the variation of entropy, $\Delta S = S ({\cal T} + \Delta t)-S({\cal T})$,
is:
\be
\Delta S \; \propto \; \left( c \Delta t \right)^2 \, + \, 
c^2 {\cal T}^2 \left( {2 \Delta t \over {\cal T}}  \right) \, . 
\ee 
The first term corresponds to the entropy of a ``small universe'', 
the universe which is ``created'', or ``opens up'' around an observer
during the time of the experiment, and embraces within its horizon the entire 
causal region about the event. The second term is a 
``cosmological'' term, that couples the local physics to the history of the 
universe. The influence of this part of the
universe does not manifest itself through elementary, classical causality
relations within the duration of the event, but indirectly, through a (slow)
time variation of physical parameters such as masses and couplings,
an effect discussed in \cite{spi} and \cite{assiom}.
In the approximation of our abstraction to the rather ideal case of
two inertial frames, we must neglect this part, concentrating the discussion
to the local physics. In this case, each experiment must be considered
as a ``universe'' in itself.
Let's indicate with $\Delta t$ the time interval as reported by $A$, and 
with $\Delta t^{\prime}$
the time interval reported by $A^{\prime}$.
In units for which $c=1$, and omitting the normalization
constant common to all the expressions like~\ref{ScT2} , 
we can therefore write:
\be
\Delta S(A) ~ = ~ ( \Delta t )^{2} \, , 
\label{deltat}
\ee
whereas
\be
\Delta S (A^{\prime}) ~ = ~ ( \Delta t^{\prime} )^{2} \, , 
\ee
and
\be
\Delta S_{\rm Kinetic} (A) ~ = ~ ( v \, \Delta t )^{2} \, . 
\label{vt2}
\ee
These expressions have the following interpretation. As seen from $A$,
the total increase of entropy corresponds to the black hole-like entropy
of a sphere of radius equivalent to the time duration of the 
experiment. Since $v=c=1$ is the maximal 
``classical'' speed of propagation of information, all the classical 
information about the system is contained within the horizon set by the 
radius $c \Delta t= \Delta t$. However, when $A$ attempts to refer this time 
measurement to what $A^{\prime}$ could observe, it knows that 
$A^{\prime}$ perceives itself at rest, and 
therefore it cannot include in the computation of entropy also the change in 
configuration due to its own motion (here it is essential that we consider 
inertial systems, i.e. constant motions). 
``$A$'' separates therefore its measurement 
into two parts, the ``internal one'', namely the one involving changes that
occur in the configuration as seen at rest by $A^{\prime}$ 
(a typical example is for instance a muon's decay at rest in
$A^{\prime}$), and a part accounting 
for the changes in the configuration due to the very being 
$A^{\prime}$ in motion at speed $v$.
If we subtract the internal changes, namely we think at the system at rest in
$A^{\prime}$ as at a point without meaningful physics 
apart from its motion in space \footnote{No internal physics means
that we also neglect the contribution to the energy/entropy due to the
mass.}, the entire information about the change of entropy is contained in 
the ``universe''
given by the sphere enclosing the region of its displacement, 
$v^2 (\Delta t)^2 ~ = ~ \Delta S_{\rm Kinetic} (A)$. In other words,
once subtracted the internal physics, the system behaves, from the point
of view of $A$, as a universe which expands at speed $v$, because the only
thing that happens is the displacement itself, of a point otherwise fixed
in the local universe (see figure~\ref{v-universe}). 
Inserting expressions \ref{deltat}--\ref{vt2} in \ref{DeltaSSk} we obtain:
\be
(\Delta t)^2 ~ = ~ { (\Delta t^{\prime})^2 \over 1 - v^2 } \, , 
\ee
that is:
\be
\Delta t ~ = ~ { \Delta t^{\prime} \over \sqrt{1 - v^2}  } \, .
\ee
The time interval as measured by $A$ results to be longer by a factor 
$(\sqrt{1 - v^2})^{-1}$ than as measured by $A^{\prime}$.
We stress that, when we use expressions like ``as seen from'', ``it observes''
and alike, we intend them in an ideal sense, not in the concrete sense of
``detecting a light ray coming from the observed''. For the derivation
of the Lorentz boost we did not make explicit use of the geometry of the
propagation of light rays between observed and observers. 
The bound on the speed of information, and therefore of light, enters 
on the other hand when we write the variation of entropy of the ``local
universe'' as $\Delta S = (c \Delta t)^2$. If $c \to \infty$, namely, if within
a finite interval of time an infinitely extended causal region opens up around
the experiment, both $A$ and $A^{\prime}$ turn out to have access to the full
information, and therefore $\Delta t = \Delta t^{\prime}$. Namely, they observe
the same overall variation of entropy.

\subsection{the space boost}

In this framework we obtain in quite a natural way the Lorentz 
\underline{time} boost. 
The reason is that, for us, time evolution is directly related to entropy 
change, and we identify configurations (and geometries) through their entropy. 
Space length is somehow a derived quantity, 
and we expect also the space boost to be 
a secondary relation. Indeed, it can be easily derived from the time boost, 
once lengths and their measurements are properly defined. However, these 
quantities are less fundamental in that they are related to the classical 
concept of geometry. 
We could produce here an argument leading to the space boost. However, 
this would basically be a copy of the classical derivation within the 
framework of special relativity. A derivation of the time boost through 
entropy-based arguments opens 
instead new perspectives, allowing to better understand where relativity ends
and quantum physics starts. Or, to better say, it provides us with an 
embedding of this problem into a scenario that contains both these aspects,
relativity and quantization, as 
particular cases, to be dealt with as useful approximations.

\section{General time coordinate transformation}
\label{gtrans}

Lorentz boosts are only a particular case of a more general
transformation. They are valid when systems are not 
accelerated; in particular, when they are not subjected to a gravitational 
force. Traditionally, we know that 
the general coordinate transformation has to 
be found within the context of General Relativity; in that case the measure 
of time lengths 
is given by the time-time component of the metric tensor. In the absence of 
mixing with space boosts, i.e., with a diagonal metric, we have:
\be
(d s)^2 ~ = ~ g_{00} (dt)^2 \, . 
\ee 
As the metric depends on the matter/energy content through the Einstein's 
Equations:
\be
{\cal R}_{\mu \nu} - {1 \over 2} g_{\mu \nu} {\cal R} \, 
= \, 8 \pi G_N T_{\mu \nu} \, ,
\ee
$g_{00}$ can be computed when we know the energy of the system. For instance,
in the case of a particle of mass $m$ moving at constant speed 
$\vec{v}$ (inertial motion),
the energy, the ``external'' energy, is the kinetic energy 
${1 \over 2} m v^2$, and we recover the
$v^2$-dependence of the Lorentz boost~\footnote{In the determination of the 
geometry, what matters here is not the full force experienced by the particle
but the field  in which the latter moves. The mass $m$ therefore drops out from
the expressions (see for instance~\cite{landau}).}.   

In the simple case of the previous section, we have considered 
the physical system of the wave packet as
decomposed into a part experiencing an ``internal'' physics, and a 
``center of mass point of view'' 
part in which the complex internal physics is dealt with as a point-like 
particle. The Lorentz boost has been derived as the consequence of a 
transformation of entropies. 
Indeed, our coordinate transformation is based on the same physical grounds
as the usual
transformation of general Relativity, based on a metric derived from the
energy tensor. In the linear
approximation, where one keeps only the first two terms of the expansion of the
square-root $\sqrt{1 - v^2/c^2}$, the Lorentz boost can be obtained from
an effective action in which in the Lagrangian appear the rest and the kinetic
energy. These terms correspond to the two terms on the r.h.s. of 
equation~\ref{DeltaSSk}. 
Entropy has in fact the dimension of an energy multiplied by a time 
\footnote{By definition, $d S = dE / T$, where $T$ is the temperature, and
remember that in the conversion of thermodynamic formulas, 
the temperature is the inverse of time.}. Approximately, we can write:
\be
\Delta S ~ \simeq ~
\Delta E \Delta t \, , 
\label{SEt}
\ee 
where $\Delta E$ is either the kinetic, or the rest energy.
The linear version of the Lorentz boost is obtained by inserting 
in~(\ref{SEt}) the expressions $\Delta E_{rest} = m$ and 
$\Delta E_{kinetic} = {1 \over 2} m v^2$. In this case, the linearization
of entropies lies in the fact that we consider the mass a constant, instead
of the full energy of the 
``local universe'' contained in a sphere of 
radius $\Delta t$, i.e. the energy (mass) of a black hole of radius 
$\Delta t$: $m = \Delta E = \Delta t / 2$. 
Although imprecise, this approach through the linear approximation
helps to understand where things come from.

In our theoretical framework, the general expression of the time coordinate
transformation is:
\be
(\Delta t^{\prime})^2 ~ = ~ 
\Delta S^{\prime}(t) \, - \, 
\Delta S^{\prime}_{external}(t) \, .
\label{SpSS}
\ee
Here $\Delta S^{\prime}(t)$ is the total variation of entropy of the ``primed''
system as measured in the ``unprimed'' system of coordinates:
$\Delta S^{\prime}(t) \, = \, (\Delta t)^2$.
We can therefore write expression~\ref{SpSS} as:
\be
(\Delta t^{\prime})^2 ~= ~ \left[ 1 - \mathcal{G}(t) \right] (\Delta t)^2 \, , 
\label{dtdt2}
\ee  
where:
\be
\mathcal{G}(t) ~ \stackrel{\mathrm{def}}{=} ~ 
{\Delta S^{\prime}_{external}(t) \over (\Delta t)^2}   \, . 
\label{Gdef}
\ee
With reference to the ordinary metric tensor $g_{\mu \nu}$, we have:
\be
\mathcal{G}(t) ~ = ~ g_{00} + 1 \, .
\ee
$\Delta S^{\prime}_{external}(t)$ is the part of change of entropy of 
$A^{\prime}$ referred to by the observer $A$
as something that does not belong to the rest frame of $A^{\prime}$. It can be
the non accelerated motion of $A^{\prime}$, as in the previous example, or more
generally the presence of an external force that produces an acceleration.
Notice that the coordinate transformation \ref{dtdt2} starts with a constant
term, $1$: this corresponds to the rest entropy term expressed in
the frame of the observer. For the observer, the new time metric is always
expressed in terms of a deviation from the identity.

By construction, \ref{Gdef} is the ratio between the metric in 
the observed system and the metric in the system of the observer.
From such a coordinate transformation we can pass to the metric of
space-time itself, provided we consider the coordinate transformation 
between the metric $g^{\prime}$ of a point in space-time,
and the metric of an observer which lies on 
a flat reference frame, whose metric is expressed in flat coordinates. 
We have then:
\be
\mathcal{G}(t) \, - \, 1 ~ = ~ { g_{00}^{(\prime)} \over g_{00} \, 
= \eta_{00} = 1} \, .
\ee
As soon as this has been clarified, we can drop out the denominator and we 
rename the primed metric as the metric tout court.

Once the measurement of lengths
is properly introduced, as derived from a measurement of configurations
along the history of the system, 
it is possible to extend the relations also to the transformation of
space lengths. This gives in general the components of the metric tensor
as functions of entropy and time. In classical terms, whenever this 
reduction is possible, this can be rephrased into a dependence on energy 
(density) and time. They give therefore a generalized, integrated version
of the Einstein's Equations.
Let's see this for what concerns the time component of the metric. We want
to show that the metric $g_{00}$ of the effective space-time corresponds
to the metric of the distribution of energy in the mean space, i.e., 
in the classical limit of effective three-dimensional space as it
arises from \ref{ZSintegral}. This will mean
that the geometry of the motion of a particle within this space is the
geometry of the energy distribution. In particular, if the energy is 
distributed according to the geometry of a sphere, so it will be the
geometry of space-time in the sense of General Relativity.
To this regard, we must remember that: 

\noindent
${\rm i})$  All this makes only sense in the ``classical
limit'' of our scenario, namely only in an average sense, where 
the universe is dominated by a configuration that can be described in 
classical geometric terms. It is in this limit that the universe
appears as three dimensional. Configurations that
are in general non three-dimensional, non-geometric, possibly tachyonic, and,
in any case, configurations for which General Relativity and Einstein 
Equations don't apply, are covered under the ``un-sharping'' relations of the
Uncertainty Principle.
All of them are collectively treated as ``quantum effects'';

\noindent
${\rm ii})$
In the classical limit, \underline{nothing}
travels at a speed higher than $c$. As during an experiment
no information comes from outside the 
local horizon set by the duration of the experiment itself, to cause
some (classical) effects on it, any consideration
about the entropy of the configuration of the object under consideration
can be made ``local'' (tachyonic effects are taken into account
by quantization). That means, when we consider the motion of an
object along space
we can just consider the local entropy, depending on, and determined by, 
the energy distribution around the object.

Having these considerations in mind,
let us consider the motion of a particle, or, more precisely, a 
non-dispersive wave-packet, in the mean, three-dimensional, classical space. 
Consider to perform a (generally pointwise) coordinate transformation to
a frame in which the metric of the energy distribution external 
to the system intrinsically building the wave packet in itself is flat,
or at least remains constant. As seen from this set of frames, along the motion
there is no change of the (local) entropy around the particle, and the right
hand side of \ref{Gdef} vanishes, implying that also the metric of the motion
itself remains constant (remember that \ref{Gdef} in
this case gives the \emph{ratio} between metrics at different points/times).
If on the other hand we keep the frame of the observer fixed,
and we ask ourselves what will be the direction chosen by the particle
in order to decide the steps of its motion, the answer will be:
the particle ``decides'' stepwise to go in the direction that maximizes
the entropy around itself. Let us consider configurations in which 
the only property of particles is their mass
(no other charges), so that entropy is directly related to the ``energy
density'' of the wave packet. In this case,  
between the choice of moving toward 
another particle, or far away, the system will proceed in order to increase
the energy density around the particle. Namely, moving 
the particle toward, rather than away from, the other particle,
in order to include in its horizon also the new system. This is how 
gravitational attraction originates in this theoretical framework.

In order to deal with more complicated cases, such as those in which
particles have properties other than just their mass 
(electro-magnetic/weak/strong charge), we need a more detailed description of 
the phase space. In principle things are the same, but the appropriate
scenario in which all these aspects are taken into account is the one
discussed in Ref.~\cite{spi}, in which these issues are phrased and addressed
within a context of (quantum) String Theory. 
In Ref.~\cite{spi} it was discussed how within that framework a full world of 
particles with masses and charges, with the correct interactions, arises.

\section{Transformations beyond the classical limit}
\label{beyondc}

Lifting the issue of relating time measurements to a problem 
of a comparison of entropies allows to address
questions that go beyond the domain of the Theory of Relativity. 
For instance, it is possible to ask
how does it appear an inertially moving frame in the
extremely relativistic regime, when the speed $v$ approaches the speed of 
light $c$, or what happens in a system in a strong gravitational field, 
or acceleration. 
According to the theory of Relativity, we would say that in the
limit $v \to c$ the time gets ``frozen''. But, if we look at
expression \ref{DeltaSSk}, we see that we are just allowed to
conclude that the variation of internal entropy decreases. Only in the
classical limit this means a slowing down of the time flow, as only in
this limit we can concentrate on the configuration corresponding
to a classical geometric description, neglecting all other configurations
contributing to the mean value of what we observe.
As discussed in Ref.~\cite{assiom}, 
out of the classical configuration, less entropic
ones contribute to what collectively we consider the ``quantum spread''
of observables.

In our scenario, a quantum system, or a system
considered in the quantum regime, is precisely a superposition of less entropic
configurations. According to Ref.~\cite{assiom}, 
these indeed correspond to the microscopic details
of our universe. This agrees with the common picture of
what is the quantum scale of phenomena \footnote{To be more precise,
in our scenario there is no exact distinction between ``classical'' and 
``quantum mechanical'', in that everything is embedded in an universal 
description, which allows in this way to
deal with ``quantum gravity aspects'' also when they manifest themselves 
on a large scale.}.
When the \emph{classical} variation of entropy becomes negligible, the system
does not freeze. In the usual language, we can say 
it starts showing up a quantum
behaviour. What happens is that, in the computation of entropy, 
we cannot anymore 
neglect the less entropic configurations. Moreover, 
we cannot go on pushing the external entropy, for instance by going to
the very edge of the limit $v \to c$, till the vanishing of the internal term,
without loosing the meaning of what we are doing. Indeed,
beyond a certain limit, the separation in two terms of the r.h.s.
of expressions \ref{DeltaSS} and \ref{DeltaSSk},
always possible from a mathematical point of view, looses its
classical sense: as we go on ``taking away'' terms from the internal
part of entropy and attributing them to the external one,
also in the external part of entropy we start to include non-classical
configurations \footnote{Not for every system 
it is easy to distinguish between 
internal and external part. There may be situations in which undergoing an 
external field of force deeply modifies the full system. 
But it is also true that in these situations
the system as it is characterized as free, ``at rest'', does not exist anymore
once in interaction. There is therefore no more ``rest frame''.}. 
In general it happens that ``peripherical'' configurations, which are
less localized, start to weight more on the superposition that builds up the
configuration of the system in its rest frame, 
and this results in an overall effective increase of its de-localization.

\vspace{2.5cm}

\providecommand{\href}[2]{#2}\begingroup\raggedright\endgroup

\end{document}